\documentclass[preprint,showpacs,preprintnumbers,amssymb,nofootinbib]{revtex4}

\usepackage{graphicx}

\begin{document}

\title{Gravitational Radiation from the radial infall of highly
relativistic point particles into Kerr black holes}

\author{Vitor Cardoso}
\email{vcardoso@fisica.ist.utl.pt}
\author{Jos\'e P. S. Lemos}
\email{lemos@kelvin.ist.utl.pt}
\affiliation{
Centro Multidisciplinar de Astrof\'{\i}sica - CENTRA, 
Departamento de F\'{\i}sica, Instituto Superior T\'ecnico,
Av. Rovisco Pais 1, 1049-001 Lisboa, Portugal
}%

\date{\today}

\begin{abstract}

In this paper, we consider the gravitational radiation generated by
the collision of highly relativistic particles with rotating Kerr
black holes.  We use the Sasaki-Nakamura formalism to compute the
waveform, energy spectra and total energy radiated during this
process.  We show that the gravitational spectrum for high-energy
collisions has definite characteristic universal features, which are
independent of the spin of the colliding objects.  We also discuss
possible connections between these results and the black hole-black
hole collision at the speed of light process.  With these results at
hand, we predict that during the high speed collision of a
non-rotating hole with a rotating one, about $35\%$ of the total
energy can get converted into gravitational waves. Thus, if one is
able to produce black holes at the Large Hadron Collider, as much as
$35\%$ of the partons' energy should be emitted during the so called
balding phase.  This energy will be missing, since we don't have
gravitational wave detectors able to measure such amplitudes.  The
collision at the speed of light between one rotating black hole and a
non-rotating one or two rotating black holes turns out to be the most
efficient gravitational wave generator in the Universe.

\end{abstract}


\maketitle
\newpage
\section{Introduction}
In previous works we have studied the collision at the speed of light
between a point particle and a Schwarzschild black hole
\cite{vitorjose1}, and a point particle and a Kerr black hole along
its symmetry axis \cite{vitorjose2}.  This analyses can describe
several phenomena, such as, the collision between small and massive
black holes \cite{schutz}, the collision between stars and massive
black holes, or the collision between highly relativistic particles
like cosmic and gamma rays colliding with black holes \cite{piran}, to
name a few.  We have argued in \cite{vitorjose1,vitorjose2}
that these studies could give valuable
quantitative answers for the collision at the speed of light between
equal mass black holes, although we only work with perturbation theory
{\it {\`a} la} Regge-Wheeler-Zerilli-Teukolsky-Sasaki-Nakamura, which
formally does not describe this process.  In fact, extrapolating for
two equal mass, Schwarzschild black holes we obtained
\cite{vitorjose1} results which were in very good agreement with
results \cite{payne} obtained through different methods.  For equal
mass black holes, we found that the collision at the speed of light
between a Schwarzschild black hole and a Kerr black hole along the
symmetry axis gave similar results to those in \cite{vitorjose1} . In
particular, we found that $15.5 \%$ of the total energy gets converted
into gravitational waves (for two non-rotating holes the amount is
slightly less, $13 \%$). There are as yet no numerical results for
rotating black holes, so there is no clue as to the correctness of our
results for rotating holes.  However, should these results give
radiated energies larger than that allowed by the cosmic censorship,
and therefore by the area theorem, we would run into problems, and we
would know that the extrapolation to equal masses is strictly
forbidden.  Our previous results do not violate the area theorem, but
the mere possibility brings us to the study of the collision of a
non-rotating black hole with a rotating one, along the equatorial
plane, both approaching each other at nearly the speed of light.
Sasaki, Nakamura and co-workers \cite{nmothers} have studied the
infall of particles, at rest at infinity into Kerr black holes, along
the equatorial plane and along the axis.  They found that the energy
radiated for radial infall along the equatorial plane was about $2.6$
times larger than for radial along the axis, for extreme holes.
Accordingly, we expect the energy for highly relativistic collisions
along the equatorial plane to be larger than that along the symmetry
axis, but how larger?  If the ratio $\frac{\Delta E_{{\rm
equator}}}{\Delta E_{{\rm axis}}}$ should still be the same, $2.6$,
then the energy released in the collision along the equatorial plane
of a Kerr hole and a non-rotating hole at the speed of light would be
$\sim 40 \%$ of the total energy.  But this is slightly more than the
upper bound on the efficiency allowed by the area theorem, which is
$\sim 38.7 \%$.  Motivated mainly by this scenario, we shall study
here the collision at nearly the speed of light between a point
particle and a Kerr black hole, along its equatorial plane.  We will
then perform a boost in the Kerr black hole, and extrapolate for two
equal mass objects, one rotating, the other non-rotating.

Another point of interest to study this process is the hypothesis that
black holes could be produced at the Large Hadron Collider (LHC) at
CERN, which has recently been put forward \cite{bhprod} in the so called
TeV-scale gravity scenarios.  In such scenarios, the hierarchy problem
is solved by postulating the existence of $n$ extra dimensions,
sub-millimeter sized, such that the $4+n$ Planck scale is equal to the
weak scale $\sim$ 1TeV. The Standard Model lives on a 4-dimensional
sub-manifold, the brane, whereas gravity propagates in all
dimensions. If TeV-scale gravity is correct, then one could
manufacture black holes at the LHC.  The ability to produce black
holes would change the status of black hole collisions from a rare
event into a human controlled one.  This calls for accurate
predictions of gravitational wave spectra and gravitational energy
emitted during black hole formation from the high speed encounter of
two particles.  This investigation was first carried out by D'Eath and
Payne \cite{payne} by doing a perturbation expansion around the
Aichelburg-Sexl metric, describing a boosted Schwarzschild black
hole. Their computation was only valid for non-rotating black holes
and it seems quite difficult to extend their methods to include for
spinning black holes.  Our approach however allows one to also study
spinning black holes, which is of great importance, since if one forms
black holes at the LHC, they will most probably be rotating ones, the
chance for having a zero impact parameter being vanishingly small.

Suppose therefore one can produce black holes at LHC. The first thing
that should happen is a release of the hole's hair, in a phase termed
``balding'' \cite{bhprod}. The total amount of energy released in such
a phase is not well known, and is based mostly in D'Eath and Payne's
results.  With our results one can predict the total gravitational
energy radiated in the balding phase when the resulting holes are
rotating, a process which, as we shall see, radiates a tremendous
amount of energy.  This means that there will be a ``missing'' energy
during the formation of a black hole, this energy being carried away
by gravitational waves, and undetected, at least by any realistic
present technology.  Our results suggest that as much as 35\% of the
center of mass energy can be leaking away as gravitational waves, and
therefore one should have 35\% missing energy.  Strictly speaking,
these values are only valid for a head-on collision, i.e. a collision
with zero impact parameter, along the equatorial plane. One expects
the total energy to decrease if the collision is taken along a
different plane, or with non-zero impact parameter.  However, we do
not expect the total energy to vary much as long as the impact parameter
stays sufficiently small enough to give rise to black hole production.
\section{The Teukolsky and Sasaki-Nakamura formalism}

In this section, we give a very brief account of the Teukolsky equation,
and of the Sasaki-Nakamura equation. Details about the Teukolsky formalism may
be found in the original literature \cite{teukolsky}, and also in 
\cite{breuerbook}. For a good account of the Sasaki-Nakamura 
formalism we refer the reader
to \cite{nmothers,nakamurasasaki,hughes1}.

We start from the Kerr background geometry, written in Boyer-Lindquist 
$(t, r, \theta, \phi)$ coordinates:

\begin{equation}
ds^{2}=-(1-\frac{2Mr}{\Sigma})dt^{2}-\frac{4Mar}{\Sigma}\sin^2\theta dt d\phi+
\frac{\Sigma}{\Delta} dr^{2}+\Sigma d\theta^{2}+
\frac{A\sin^2\theta}{\Sigma} d\phi^{2}\,,
\label{lineelementKerrBoyer}
\end{equation}
Here, $M$ is the mass of the black hole, and $a$ its angular momentum per
unit mass.
Also $\Sigma= r^2+a^2\cos^2\theta \,$ ; $\Delta=r^2+a^2-2Mr \,$ ; 
$A=(r^2+a^2)^2-\Delta a^2\sin^2\theta$.

Working with Kinnersley's null tetrad, one can show \cite{teukolsky}
that the equations for the Newman-Penrose quantities decouple and
separate, giving rise to the Teukolsky equation,

\begin{equation}
\Delta^{-s}\frac{d}{dr}(\Delta^{s+1}\frac{d}{dr} {_sR})-\, _sV \, _sR 
=-\, _sT \,.
\label{teukolskyequation}
\end{equation}
Here,
\begin{eqnarray}
 _sV=\frac{-K^2}{\Delta}+
\frac{isK \Delta'}{\Delta}-\frac{2isK'}{\Delta}-2isK'+\, _s\lambda \\
K=(r^2+a^2)\omega - am \,,
\label{teukolskyequationexplanation}
\end{eqnarray}
and a prime denotes derivative with respect to $r$.
The quantity $s$ denotes the spin weight (or helicity) 
of the field under consideration
(see for example \cite{newman}) and $m$ is an azimuthal quantum number. 
We are interested in gravitational perturbations, which have spin-weight
$s=-2,+2$. Solutions with $s=-2$ are related to those with $s=2$ via
the Teukolsky-Starobinsky identities, so we need only worry about a
specific one. For definiteness, and because that was the choice adopted
by Sasaki and others \cite{nmothers}, we work with $s=-2$.  The constants
$m$, $_s\lambda$ ($_s\lambda$ depends non-linearly on $\omega$) are
separation constants arising from the azimuthal function $e^{im\phi}$
and the angular eigenfunction $\,_sZ_{lm}^{a\omega}(\theta,\phi)$,
respectively. 
$\,_sZ_{lm}^{a\omega}(\theta,\phi)$ is the spin-weighted spheroidal
harmonic,
$_sZ_{lm}^{a\omega}(\theta,\phi)=
\frac{1}{(2\pi)^{1/2}}\,_sS_{lm}^{a\omega}(\theta)e^{im\phi}$, 
and in turn, $_sS_{lm}^{a\omega}(\theta)$ satisfies
\begin{equation}
\frac{1}{\sin\theta}\frac{d}{d\theta}(\sin \theta \frac{d}{d\theta} {S})+
(a^2\cos^2\theta-\frac{m^2}{\sin^2\theta}+4a
\omega\cos\theta+\frac{4m\cos\theta}{\sin^2\theta}
-4\cot^2\theta+c)S=0\,, 
\label{spinspheroidal}
\end{equation}
where $c=\lambda-2-a^2\omega^2+2am\omega$. The spheroidal functions
are normalized according to $\int d\Omega
|_sZ_{lm}^{a\omega}(\theta,\phi)|^2=1$.  These functions have not been
thoroughly exploited as far as we know, but the essentials, together
with tables for the eigenvalues $\lambda$ for $s=-2$ can be found in
\cite{pressteu,breuer}.  The source term $_sT$ appearing in
(\ref{teukolskyequation}) may be found in \cite{breuerbook}, and
depends on the stress-energy tensor of the perturbation under
consideration.  We note that, for a test particle falling in from
infinity with zero velocity, for example, $T \sim r^{7/2}$ and $V$ is
always long-range. This means that when we try to solve the Teukolsky
equation (\ref{teukolskyequation}) numerically we will run into
problems, since usually the numerical solution is accomplished by
integrating the source term times certain homogeneous solutions. At
infinity this integral will not be well defined since the source term
explodes there, and also because since the potential is long range,
the homogeneous solution will also explode at infinity.  We can remedy
this long-range nature of the potential and at the same time
regularize the source term in the Sasaki-Nakamura formalism. After a set of
transformations, the Teukolsky equation (\ref{teukolskyequation}) may
be brought to the Sasaki-Nakamura form
\cite{nakamurasasaki}:
\begin{equation}
\frac{d^2}{dr_*^2} {X(\omega,r)}- {\cal F}\frac{d}{dr_*}{X(\omega,r)} - 
{\cal U} X(\omega,r) = {\cal L} \,.
\label{sn}
\end{equation}
The functions ${\cal F}$ and ${\cal U}$ can be found in the original
literature \cite{nmothers,nakamurasasaki}. 
The source term ${\cal L}$ is given by
\begin{equation}
{\cal L}=\frac{\gamma_0\Delta}{r^2(r^2+a^2)^{3/2}}W e^{-i\int K/\Delta dr_*}\,,
\label{sourceterm}
\end{equation}
and $W$ satisfies
\begin{equation}
W''=-\frac{r^2}{\Delta}T e^{i\int K/\Delta dr_*}.
\label{W}
\end{equation}
The tortoise $r_*$ coordinate is defined by
$dr_*/dr=(r^2+a^2)/\Delta$, and ranges from $-\infty$ at the horizon
to $+\infty$ at spatial infinity.
The Sasaki-Nakamura equation (\ref{sn}) is to be solved under the
``only outgoing radiation at infinity'' boundary condition (see
section 4.2), meaning
\begin{equation}
X(\omega,r)=X^{{\rm out}}e^{i\omega r_*} \,, r_* \rightarrow \infty.
\label{xout}
\end{equation}
The two independent polarization modes of the metric, 
$h_+$ and $h_{\times}$, are given by
\begin{equation}
h_+ + ih_{\times}=\frac{8}{r\sqrt{2\pi}}\int_{-\infty}^{+\infty}d\omega 
\sum_{l,m} e^{i\omega(r_*-t)}   
\frac{X^{{\rm out}}S_{lm}^{a\omega}(\theta)}{\lambda(\lambda+2)-
12i\omega -12a^2\omega^2}e^{im\phi},
\label{definitionh}
\end{equation}
and the power per frequency per unit solid angle is
\begin{equation}
\frac{d^2E}{d\Omega d\omega}=
\frac{4\omega^2}{\pi}|\sum_{l,m}\frac{X^{{\rm out}}}{\lambda(\lambda+2)-
12i\omega -12a^2\omega^2} 
S_{lm}^{a\omega}|^2.
\label{power}
\end{equation}
Following Nakamura and Sasaki \cite{nmothers} the multipolar 
structure is given by 
\begin{equation}
h_+ + ih_{\times}=-\frac{(2\pi)^{1/2}}{r}\int_{-\infty}^{+\infty}d\omega 
\sum_{l,m}e^{i\omega(r_*-t)}\left[h^{lm}(\omega) _{-2}Y_{lm}\right]e^{im\phi},
\label{definitionhmult}
\end{equation}
which defines $h^{lm}$ implicitly. Moreover, 
\begin{equation}
\Delta E=\frac{\pi}{4}\int_{0}^{\infty}\sum_{lm}|h^{lm}(\omega)|^2+
|h^{lm}(-\omega)|^2.
\label{power2}
\end{equation}
Here, $_{-2}Y_{lm}(\theta)$ are spin-weighted spherical harmonics (basically
$_{2}S_{lm}^{a\omega=0}(\theta)$), whose properties are described by 
Newman and Penrose \cite{newman} and Goldberg et al 
\cite{goldberg}.

For the befenit of comparison, 
the general relation between Sasaki and Nakamura's $X(\omega,r)$ function
and Teukolsky's $R(\omega,r)$ function is given in \cite{nmothers} is 
terms of some complicated expressions. One finds however that near infinity, which is
the region we are interested in, this relation simplifies enormously.
In fact, when $r_* \rightarrow \infty$, the relation is
\begin{equation}
R= R^{{\rm out}} r^3 e^{i\omega r_*}\, 
\label{relationRX1}
\end{equation}
where, 
\begin{equation}
R^{{\rm out}}\equiv
-\frac{4\omega ^2 X^{{\rm out}}}{\lambda(\lambda+2)-12i
\omega -12a^2\omega^2}\,.
\label{relationRX2}
\end{equation}

\section{The Sasaki-Nakamura equations for a point particle
falling along a geodesic of the Kerr black hole spacetime}
The dependence of the emitted gravitational wave power on the
trajectory of the particle comes entirely from the source ${\cal L}$
in eqs. (\ref{sn})-(\ref{sourceterm}), which in turn through (\ref{W})
depends on the source term $T$ in Teukolsky's equation
(\ref{teukolskyequation}).  To compute the explicit dependence, one
would have to consider the general expression for the geodesics in the
Kerr geometry \cite{chandra}, which in general gives an end result not
very amenable to work with. However we find that by working in the
highly relativistic regime, the equations simplify enormously, and in
particular for radial infall (along the symmetry axis or along the
equator) it is possible to find a closed analytic and simple
expression for $W$ in (\ref{sourceterm})-(\ref{W}).  Accordingly, we
shall in the following give the closed expressions for ${\cal L}$ in
these two special situations: radiall infall along the symmetry axis
and radial infall along the equator of an highly energetic particle.

\subsection{The Sasaki-Nakamura equations for an highly relativistic
point particle falling along the symmetry axis of the Kerr black hole}

We now specialize then all the previous equations to the case under study.
We suppose the particle to be falling radially along the 
symmetry axis of the Kerr hole,
in which case all the equations simplify enormously. 
In this case the geodesics, written in Boyer-Lindquist coordinates, are
\begin{eqnarray}
\frac{dt}{d\tau}= \frac{\epsilon_0(r^2+a^2)}{\Delta}\,;\quad
\,(\frac{dr}{d\tau})^2=\epsilon_0^2-\frac{\Delta}{r^2+a^2}\,;\quad
\frac{d\phi}{d\tau}= \frac{a\epsilon_0}{\Delta}\,,
\label{geodesics1}
\end{eqnarray}
where the parameter $\epsilon_0$ is the energy
per unit rest mass of the infalling particle.  On considering highly
relativistic particles, $\epsilon_0 \rightarrow \infty$, we find that
the term ${\cal L}$ takes the simple form (compare with the source
term in \cite{vitorjose1,vitorjose2}) :
\begin{equation}
{\cal L}=-\frac{\mu C \epsilon_0 \gamma_0 \Delta}{2\omega^2 
r^2 (r^2+a^2)^{3/2}}e^{-i\omega r_*}.
\label{explicitS1}
\end{equation}
Here,
$C=\left[\frac{8Z_{l0}^{a\omega}}{\sin^2\theta}\right]_{\theta=0}$,
the function $\gamma_0=\gamma_0(r)$ can be found in
\cite{nmothers}, while $\mu$ is the mass of the particle.

\subsection{The Sasaki-Nakamura equations for an highly relativistic
point particle falling along the equatorial plane of the Kerr black
hole}

We now suppose the particle is falling radially along the
equator of the Kerr hole, where radially is defined in the sense of
Chandrasekhar \cite{chandra}, meaning that $L=a\epsilon_0$, where $L$
is the conserved angular momentum. In this case the geodesics, written
in Boyer-Lindquist coordinates, are
\begin{eqnarray}
\frac{dt}{d\tau}= \frac{\epsilon_0(r^2+a^2)}{\Delta}\,;\quad
\,r^2(\frac{dr}{d\tau})^2=\epsilon_0^2r^2-\Delta\,;\quad
\frac{d\phi}{d\tau}= \frac{a\epsilon_0}{\Delta}\,.
\label{geodesics2}
\end{eqnarray}
On considering highly relativistic particles, $\epsilon_0 \rightarrow
\infty$, we find that the source term takes again a very simple form
(compare with the source term in \cite{vitorjose1,vitorjose2}):
\begin{equation}
{\cal L}=-\frac{\mu \hat{S}\epsilon_0 \gamma_0 
\Delta}{\omega^2 r^2 (r^2+a^2)^{3/2}}
e^{-i\int\frac{K}{\Delta}dr}\,,
\label{explicitS2}
\end{equation}
where
\begin{equation}
\hat{S}=\left[\frac{\lambda}{2}-(m-a\omega)^2\right]Z_{lm}
({\theta=\frac{\pi}{2}})
+\left[(m-a\omega)\right]Z'_{lm}({\theta=\frac{\pi}{2}})
\label{termofontespin}
\end{equation}

Once $X^{{\rm out}}$ is known, we can get the Teukolsky 
wavefunction near infinity 
from (\ref{relationRX1}), and the waveform and energy spectra from 
(\ref{definitionh})-(\ref{power2}).

\section{Implementing a numerical solution}
To compute the spin-weighted spheroidal functions, one would have to
compute the eigenvalues $\lambda$, for each $\omega$, using a shooting
method \cite{numrecipes} or any other numerical scheme one chooses
\cite{nmothers,hughes1}.  We have chosen to lean ourselves on work
already done by Press and Teukolsky \cite{pressteu} and to verify at
each step the correctness of their results (we found no
discrepancies).  Press and Teukolsky have found, for each ($l$, $m$),
a sixth-order polynomial (in $a \omega$) which approximates the
eigenvalue $\lambda$, to whithin five decimal places (see their Table
1).  This is a good approximation as long as $a\omega<3$, which is
within our demands for $l<6$.  The computation of the spheroidal
function then follows trivially, by numerically integrating
(\ref{spinspheroidal}).

The next step is to determine $X(\omega,r)$ from the Sasaki-Nakamura
differential equation (\ref{sn}). This is accomplished by a Green's
function technique, constructed so as to satisfy the usual boundary
conditions, i.e., only ingoing waves at the horizon ($X \sim
e^{-i\omega r_*}\,,r_*\rightarrow -\infty$) and outgoing waves at
infinity($X \sim e^{i\omega r_*}\,,r_*\rightarrow \infty$).  We get
that, near infinity, and for infall along the equator (we are
interested in knowing the wavefunction in this region),
\begin{equation}
X=X^{\infty}\int \frac{X^{H}{\cal S}}{{\rm Wr}} dr_*\,;\quad
X^{\rm out}=\int \frac{X^{H}{\cal S}}{{\rm Wr}} dr_*.
\label{solution}
\end{equation}
The similar expression for infall along the symmetry axis is given in 
\cite{vitorjose2}.
Here, $X^{\infty}$ and $X^{H}$ are two linearly independent 
homogeneous solutions of (\ref{sn}) which
asymptotically behave as
\begin{eqnarray}
X^{H} \sim A(\omega)e^{i\omega r_*}+B(\omega)e^{-i\omega r_*}\,,r_* 
\rightarrow \infty \label{behavior1}\\
X^{H} \sim e^{-i\omega r_*}\,,r_* \rightarrow -\infty \label{behavior12}\\ 
X^{\infty} \sim e^{i\omega r_*}\,,r_* \rightarrow \infty 
\label{behavior13}\\
X^{\infty} \sim C(\omega)e^{i\omega r_*}+D(\omega)e^{-i\omega r_*}
\,,r_* \rightarrow -\infty\,, 
\label{behavior14}
\end{eqnarray}
and ${\rm Wr}$ is the wronskian of these two solutions.  Expression
(\ref{solution}) can be further simplified in the case of infall
along the equator to
\begin{equation}
X^{{\rm out}}=-\frac{\mu \epsilon_0 c_0 \hat{S} }{2i\omega^3 B}\int 
\frac{e^{-i\int \frac{K}{\Delta}dr} X^{H}}{r^2(r^2+a^2)^{1/2}} dr.
\label{solution2}
\end{equation}

To implement a numerical solution, we first have to determine $B(\omega)$.
We find $B(\omega)$ by solving (\ref{sn}) with the right hand side set 
to zero, and with the
starting condition $X^H= e^{-i\omega r_*}$ imposed at a large negative
value of $r_*$. For computational purposes good accuracy is hard to
achieve with the form (\ref{behavior1}), so we used an asymptotic
solution one order higher in $1/(\omega r)$:
\begin{eqnarray}
&X^H = A(\omega)(1+\frac{i(\lambda+2+2am\omega -2i\omega P)}
{2\omega r})e^{i\omega r_*}+
\nonumber\\ &
+B(\omega)(1-\frac{i(\lambda+2+2am\omega -2i\omega P)}
{2 \omega r})e^{-i\omega r_*}\,,r_* \rightarrow +\infty\,.&
\label{behavior2}
\end{eqnarray}
Here, $P=\frac{8ai(-m\lambda +a\omega(\lambda-3))}
{12a^2\omega^2+12iM\omega-\lambda^2-2\lambda-12am\omega}$.
These expressions all refer to the case of infall along 
the equator. For analogous
expressions for infall along the symmetry axis see \cite{vitorjose2}. 

In the numerical work, we chose to adopt $r$ as the independent
variable, thereby avoiding the numerical inversion of $r_*(r)$.
A fourth order Runge-Kutta routine started the integration
of $X^H$ near the horizon, at $r=r_+ + r_+\epsilon$
(the horizon radius $r_+$ is $r_+=M+(M^2-a^2)^{1/2}$), with tipically 
$\epsilon=10^{-5}$. It then integrated out to large values
of $r$, where one matches $X^H$ extracted numerically with the
asymptotic solution (\ref{behavior2}), in order to find $B(\omega)$.
To find $\Delta E$ the integral in (\ref{power2}) is
done by Simpson's rule. For both routines Richardson
extrapolation is used.

\section{Numerical Results and Conclusions}

Recent studies \cite{vitorjose1,vitorjose2} on high energy
collisions of point particles with black holes point to the existence
of some characteristic features of these processes, namely: (i) the
spectrum and waveform largely depend upon the lowest quasinormal
frequency of the spacetime under consideration;  (ii) there is a
non-vanishing zero frequency limit (ZFL) for the spectra,
$\frac{dE}{d\omega}_{\omega\rightarrow 0}$, and it seems to be
independent of the spin of the colliding particles whereas for
low-energy collisions the ZFL is zero;  (iii) the energy radiated in
each multipole has a power-law dependence rather than exponential for
low-energy collisions.

The present study reinforces all these aspects.  In Fig. 1 we show,
for an almost extreme Kerr hole with $a=0.999M$, the energy spectra
for the four lowest values of $l$, when the particle collides along
the equatorial plane.  
In Fig. 2 we show the same values but for a
collision along the symmetry axis.  The existence of a non-vanishing
ZFL is evident, but the most important in this regard is that the ZFL
is exactly the same, whether the black hole is spinning or not, or
whether the particle is falling along the equator or along the
symmetry axis.  
In fact, our numerical results show that, up to the
numerical error of about $1\%$ the ZFL is given by Table 1 
(see also \cite{vitorjose1} and the exact value given by Smarr \cite{smarr2}),
and this holds for highly relativistic particles falling along the
equator (present work), along the symmetry axis \cite{vitorjose2}, or
falling into a Schwarzschild black hole \cite{vitorjose1}.

\begin{figure}
\centerline{\includegraphics[width=10 cm,height=5 cm]
{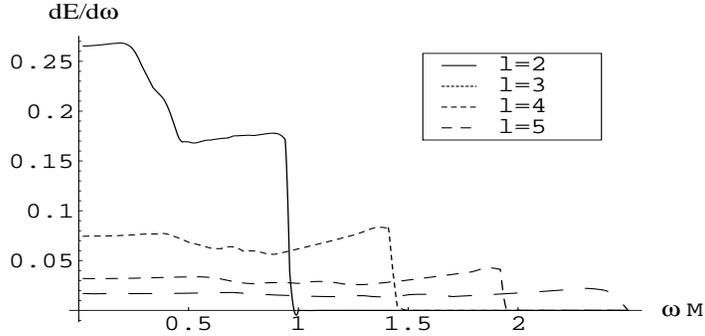}}
\caption{The energy spectra for a point particle moving at nearly the
speed of light and colliding, along the equatorial plane, with an
extreme (a=0.999M) Kerr black hole. The energy is normalized in units
of $\mu^2 \epsilon_0^2$.  Notice that the spectra is almost flat (for
large $l$), the ZFL is non-vanishing and that the quadrupole carries
less than half of the total radiated energy.}
\label{fig:1}
\end{figure}
\begin{figure}
\centerline{\includegraphics[width=10 cm,height=5 cm]
{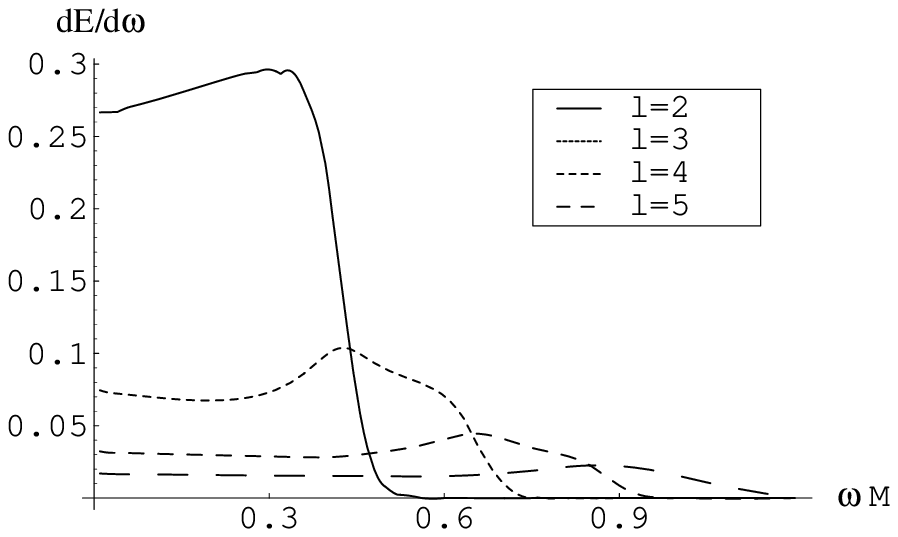}}
\caption{The energy spectra for a point particle moving at nearly the
speed of light and colliding, along the symmetry axis, with an extreme
(a=0.999M) Kerr black hole. The energy is normalized in units of
$\mu^2 \epsilon_0^2$.  Notice that the spectra is almost flat (for
large $l$), the ZFL is non-vanishing and that the quadrupole carries
less than half of the total radiated energy.}
\label{fig:2}
\end{figure}

\begin{table}
\caption{\label{tab:zfl}  The zero frequency limit (ZFL) 
for the ten lowest radiatable multipoles.}
\begin{ruledtabular}
\begin{tabular}{llll}  \hline
$l$ & ZFL($\times\frac{1}{\mu^2 \epsilon_0^2}$)&$l$& ZFL($\times 
\frac{1}{\mu^2 \epsilon_0^2}$)\\ \hline
2    &  0.265 &  7 &  0.0068 \\ \hline 
3    &  0.075 &   8 &  0.0043  \\ \hline 
4    &  0.032 &   9 &  0.003 \\ \hline 
5    &  0.0166 &   10 &  0.0023  \\ \hline 
6    &  0.01  &   11 &  0.0017 \\ \hline 
\end{tabular}
\end{ruledtabular}
\end{table}
\vskip 1mm

The $l$-dependence of the energy radiated is a power-law; in fact for large $l$
we find for infall along the equator
\begin{equation}
\Delta E_l=0.61\frac{\mu^2 \epsilon_0^2}{M l^{1.666}}\;\;,\quad\quad a=0.999M.
\label{lbehav}
\end{equation} 
Such a power-law dependence seems to be universal for high energy
collisions.  Together with the universality of the ZFL this is one of
the most important results borne out of our numerical studies.  The
exponent of $l$ in (\ref{lbehav}) depends on the rotation
parameter. As $a$ decreases, the exponent increases monotonically, until
it reaches the Schwarzschild value of 2 ($\Delta E_l \sim
\frac{1}{l^{2}}$) which was also found for particles falling along the
symmetry axis of a Kerr hole. In Table 2 we show the values of the
exponent, as well as the total energy radiated, for some values of the
rotation parameter $a$.
\vskip 1cm
\begin{center}
\begin{tabular}{|l|l|l|l|}  \hline 
\multicolumn{4}{|c|}{ $\Delta E_l=cl^{-b}$} \\ \hline
$\frac{a}{M}$&  $c$  &  $b$  & $\Delta E_{{\rm tot}}\frac{M}{\mu^2 \epsilon_0^2}$ \\ \hline
0.999        &  0.61 & 1.666 & 0.69   \\ \hline
0.8          &  0.446& 1.856 & 0.36   \\ \hline
0.5          & 0.375 & 1.88  & 0.29   \\ \hline
0            & 0.4   & 2     & 0.26   \\ \hline
\end{tabular}
\end{center}
\vskip 1mm
{\noindent Table 2. Power-law dependence of the energy radiated in
each multipole $l$, here shown for some values of $a$, the rotation
parameter. We write $\Delta E_l=\frac{c}{l^b}$ for the energy emitted
for each $l$ and $\Delta E_{{\rm tot}}$ for the total energy radiated
away. The collision happens along the equatorial plane.}
\vskip 8mm
This power-law dependence and our numerical results allow us to infer
that the total energy radiated for a collision along the equator is
\begin{equation}
\Delta E_{{\rm tot}}=0.69 \frac{ \mu^2 \epsilon_0^2}{M}\;\;,\quad\quad 
a=0.999M.
\label{Etot}
\end{equation} 
This represents a considerable enhancement of the total radiated
energy, in relation to the Schwarzschild case \cite{vitorjose1} or
even to the infall along the symmetry axis \cite{vitorjose2}.  Again,
the energy carried by the $l=2$ mode ($\Delta E_{l=2}=0.2\frac{\mu^2
\epsilon_0^2}{M}\;,\;\; a=0.999M$) is much less than the total radiated
energy.  
\begin{figure}
\centerline{\includegraphics[width=10 cm,height=5 cm]
{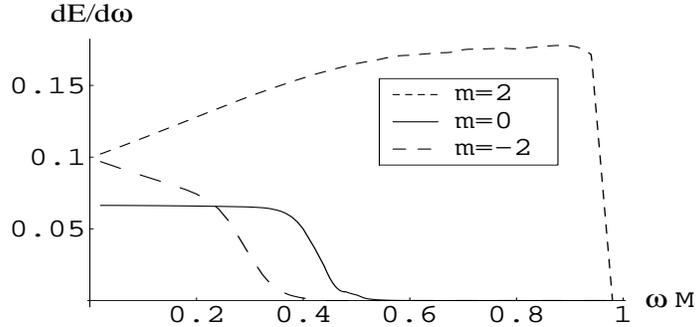}}
\caption{The energy spectra as a function of $m$ for $l=2$ 
and for an highly
relativistic particle falling along the equatorial plane 
of a Kerr black hole.
The energy is normalized in units of $\mu^2 \epsilon_0^2$.}
\label{fig:3}
\end{figure}
\begin{figure}
\centerline{\includegraphics[width=10 cm,height=5 cm]
{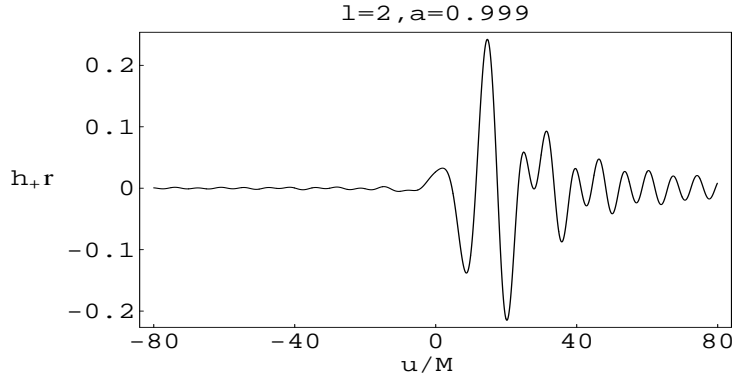}}
\caption{The $l=2$ waveform as defined by (\ref{definitionhmult}) 
for an highly
relativistic particle falling along the equatorial plane of 
an extreme
(a=0.999) Kerr black hole.The waveform is normalized in 
units of $\mu \epsilon_0$.}
\label{fig:4}
\end{figure}
In Fig. 3 we show, for $l=2$, the energy spectra as a
function of $m$, which allows us to see clearly the influence of the
quasinormal modes.  Indeed, one can see that the energy radiated for
$m=2$ is much higher than for $m=-2$, and this is due to the behaviour
of the quasinormal frequency for different azimuthal numbers $m$
\cite{QNKerr}, as emphasized by different authors \cite{nmothers}.  A
peculiar aspect is that the energy radiated for $m+l=$ odd is about
two orders of magnitude lower than for $m+l=$ even, and that is why
the spectra for $m=-1,1$ is not plotted in Fig. 3.  Notice that the ZFL
is the same for $m=2$ and for $m=-2$ both conspiring to make the ZFL
universal.  In Fig. 4 we show the $l=2$ waveform for $a=0.999M$ as
seen in $\phi=0\,, \theta=\pi /2$.  By symmetry, $h_{\times}=0$.

Up to now, we dealt only with the head-on collision between a particle
and a black hole. What can we say about collisions with non-zero
impact parameter, and along the equator ?  As is clear from our
results, the fact that the quasinormal modes are excited is
fundamental in obtaining those high values for the total energy
radiated.  Previous studies \cite{QNKerr} show that the quasinormal
modes are still strongly excited if the impact parameter is less than
$2M$. We are therefore tempted to speculate that as long as the impact
parameter is less than $2M$ the total energy radiated is still given
by Table 2. For larger values of the impact parameter, one expects
that total energy to decrease rapidly.  On the other hand, If the
collision is not along the equatorial plane, we do expect the total
energy to decrease. For example, if the collision is along the
symmetry axis, we know \cite{vitorjose2} that the total energy is
$0.31 \frac{ \mu^2 \epsilon_0^2}{M}$ for $a=0.999M$.  So we expect
that as the angle between the collision axis and the equator is varied
between $0$ and $\pi/2$ the total energy will be a monotonic function
decreasing from $0.69 \frac{ \mu^2 \epsilon_0^2}{M}$ to $0.31 \frac{
\mu^2 \epsilon_0^2}{M}$.  Still, more work is necessary to confirm
this.  Let us now consider, using these results, the collision at
nearly the speed of light between a Schwarzschild and a Kerr black
hole, along the equatorial plane.  We have argued in previous papers
\cite{vitorjose1,vitorjose2} that the naive extrapolation $\mu
\rightarrow M$ may give sensible results, so let's pursue that idea
here.  We obtain an efficiency of $34.5 \%$ for gravitational wave
generation, a remarkable increase relative to the
Schwarzschild-Schwarzschild collision.  Now, the area theorem gives an
upper limit of $38.7 \%$ so we may conclude with two remarks: (a)
these perturbative methods pass the area theorem test; (b) we are
facing the most energetic events in the Universe, with the amazing
fraction of $34.5 \%$ of the rest mass being converted into
gravitational waves.
\section*{Acknowledgements}
V. Cardoso acknowledges useful correspondence with Kin-ya Oda.
This work was partially funded by Funda\c c\~ao para a
Ci\^encia e Tecnologia (FCT) -- Portugal through project PESO/PRO/2000/4014. V.C.
also acknowledges finantial support from FCT through PRAXIS XXI
programme.  J. P. S. L. thanks Observat\'orio Nacional do Rio de
Janeiro for hospitality.

%

\end{document}